\documentclass[journal]{IEEEtran}
\usepackage[edges]{forest}
\usetikzlibrary{arrows.meta}
\usepackage{graphicx,times,amsmath,cite,enumerate}
\usepackage{kantlipsum}
\usepackage{threeparttable}
\usepackage{epsfig}
\usepackage{amssymb}
\usepackage{latexsym}
\usepackage{psfrag}
\usepackage{setspace}
\usepackage{color}
\usepackage{verbatim}
\usepackage{amsthm}
\usepackage{footnote}
\usepackage{textcase}
\usepackage{array}
\newcolumntype{P}[1]{>{\centering\arraybackslash}p{#1}}
\usepackage{float} 
\usepackage{multirow}
\usepackage{textcomp}
\usepackage{forest}
\PassOptionsToPackage{hyphens}{url}
\usepackage{url}
\usepackage{physics}
\usepackage{hyperref}
\usepackage{etoolbox}
\usepackage[edges]{forest}
\usepackage[ruled,vlined]{algorithm2e}
\usepackage{siunitx}

\usepackage{subfigure}
\usepackage{physics}
\usepackage{tikz}
\usetikzlibrary{automata, positioning}
\usepackage{nomencl}
\usepackage{float}
\usepackage{tabularx,booktabs}
\usepackage{lipsum}

\title{\vspace{-10pt}Weather Sensitive High Spatio-Temporal Resolution Transportation Electric Load Profiles For Multiple Decarbonization Pathways \vspace{-0.15em}}

\IEEEoverridecommandlockouts

\begin{document}
\bstctlcite{IEEEexample:BSTcontrol} 

\author{
\IEEEauthorblockN{Samrat Acharya\IEEEauthorrefmark{1}, Malini Ghosal\IEEEauthorrefmark{1}, Travis Thurber\IEEEauthorrefmark{1}, Casey D. Burleyson\IEEEauthorrefmark{1}, Yang Ou\IEEEauthorrefmark{2}, Allison Campbell\IEEEauthorrefmark{1}, Gokul Iyer\IEEEauthorrefmark{2}, Nathalie Voisin\IEEEauthorrefmark{1}\IEEEauthorrefmark{3}, and Jason Fuller\IEEEauthorrefmark{1} }\\
\IEEEauthorblockA{\IEEEauthorrefmark{1} \textit{Pacific Northwest National Laboratory, Richland, WA 99354, USA}} \\
\IEEEauthorblockA{\IEEEauthorrefmark{2} \textit{Joint Global Change Research Institute, Pacific Northwest National Laboratory, College Park, MD 20740, USA}} \\
\IEEEauthorblockA{\IEEEauthorrefmark{3} \textit{University of Washington, Seattle, WA 98195, USA}} \\
\{samrat.acharya, malini.ghosal\}@pnnl.gov \vspace{-10pt}
}

\maketitle
\begin{abstract}
Electrification of transport compounded with climate change will transform hourly load profiles and their response to weather. Power system operators and EV charging stakeholders require such high-resolution load profiles for their planning studies. However, such profiles accounting whole transportation sector is lacking. Thus, we present a novel approach to generating hourly electric load profiles that considers charging strategies and evolving sensitivity to temperature. The approach consists of downscaling annual state-scale sectoral load projections from the multi-sectoral Global Change Analysis Model (GCAM) into hourly electric load profiles leveraging high resolution climate and population datasets. Profiles are developed and evaluated at the Balancing Authority scale, with a 5-year increment until 2050 over the Western U.S. Interconnect for multiple decarbonization pathways and climate scenarios. The datasets are readily available for production cost model analysis. Our open source approach is transferable to other regions.


\end{abstract}

\section{Introduction}
\label{sec:intro}

The transportation sector is a major greenhouse gas (GHG) emitter, accounting for 29\% of United States (U.S.) \cite{us_epa} and 27\% of global emissions in 2019 \cite{iea_ghg_sector}. It is the largest GHG emitter in the U.S. and second largest globally. While the focus has primarily been on electrifying light-duty vehicles (LDVs), the electrification of medium- and heavy-duty vehicles (MHDVs) and non-road vehicles (such as trains, aviation, and ships) is gaining momentum \cite{mai2018electrification}. Currently, LDVs have seen significant electrification, with global electric car sales reaching 16.5 million in 2021, representing 9\% of the global car market \cite{IEA}. Electric bus and truck penetration is lower at 4\% and 0.1\% respectively \cite{IEA}. However, achieving comprehensive decarbonization requires electrifying the entire transportation sector, including MHDVs and non-road vehicles. 


Rapid large-scale transportation electrification poses challenges to economics \cite{kintner2020electric}, operations \cite{lopes2010integration}, and cybersecurity \cite{acharya2020public} of power grids and EV chargers. To address these challenges, accurate and highly granular projections of spatio-temporal transportation charging load profiles are essential for effective planning and decision-making.
Recent studies \cite{gaete2021open, borlaug2021heavy, osti_1836645} have employed data-driven approaches to project EV charging load profiles. For example, Gaete et al. \cite{gaete2021open} developed LDV charging profiles in Germany using data on mobility, battery characteristics, and charging strategies. 

\begin{table}[H]
\centering
\begin{tabular}{ |p{0.46\textwidth}| } 
 \hline
 This paper is accepted for publication in 2024 IEEE ISGT NA. The complete copyright version will be available on IEEE Xplore when the conference proceedings are published. \\
 \hline
\end{tabular}
\end{table}

Borlaug et al. \cite{borlaug2021heavy} created average daily charging profiles for depot-based heavy-duty trucks based on truck mobility and charging data. Wang et al. \cite{osti_1836645} generated daily charging profiles for MHDVs in California by considering mobility patterns and future charger capacities. In contrast to these studies \cite{gaete2021open, borlaug2021heavy, osti_1836645}, our approach encompasses the entire transportation sector, providing comprehensive time-series data for charging load profiles.

\begin{figure}[!t]
    \centering
    \vspace{-3pt}
\includegraphics[width=0.99\columnwidth, clip=true, trim= 5mm 55mm 10mm 5mm]{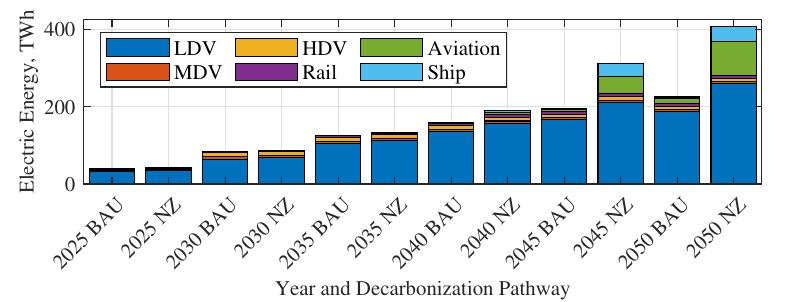}
    \caption{Energy use of LDVs, MDVs, HDVs, and non-road vehicles in the western U.S. interconnection for the Business As Usual (BAU) and Net Zero (NZ) decarbonization pathways in  the Global Change Analysis Model.}
    \label{fig:penetration}
\end{figure}

In this paper we analyze transportation electrification in the western U.S. interconnection region. Our analysis encompasses two decarbonization pathways: i) Business As Usual (BAU) and ii) Net-Zero (NZ), modeled using the U.S. version of the Global Change Analysis Model (GCAM-USA) as depicted in Fig.~\ref{fig:penetration}. GCAM is a widely utilized, economically-driven, multi-sector dynamics model that simulates the interactions between human and natural systems, playing a crucial role in assessing global changes and their impacts. Notably, GCAM has been employed by organizations like the Intergovernmental Panel on Climate Change (IPCC) \cite{ipcc} and the U.S. Department of Energy. The BAU scenario in GCAM reflects existing GHG emission reduction policies. The NZ scenario embodies ambitious plans to achieve a clean U.S. grid by 2035 and net-zero GHG emissions by 2050. Leveraging these decarbonization pathways, we generate hourly transportation electric load profiles across Balancing Authorities (BAs) in the western U.S. interconnection. BAs oversee resource planning, electricity supply-demand balance, and real-time frequency regulation in specific geographic areas. For spatial details on BAs in the western U.S. interconnection, we refer to \cite{ba_map}. To this end, we make the following unique contributions:

\begin{enumerate}
\item We develop a spatially-distributed statistical downscaling approach, leveraging multi-sector dynamics in GCAM, to project annual transportation energy onto hourly BA-level time-series profiles in the western U.S. interconnection. 

\item We conduct a detailed sensitivity analysis of transportation charging profiles considering climate change scenarios, decarbonization pathways, and charging strategies.
\item We provide open source code at \url{https://doi.org/10.5281/zenodo.7888569}\cite{acharya_samrat_2023_7888569} to accelerate the dissemination of the approach and support the community of practice.
\end{enumerate}

\section{Data and Tools}
\label{sec:methodology}
This section outlines the key data and tools employed to project spatio-temporal transportation charging load profiles.

\subsection{Global Change Analysis Model (GCAM)}
\label{sec:gcam}


The Global Change Analysis Model (GCAM) \cite{GCAMv6documentation} is an open-source integrated assessment model that encompasses energy, economy, agriculture and land use, water, and climate systems at global spatial scales. For this study GCAM spans from 2015 to 2100, with 5-year time-steps, and the final calibration year is 2015. We utilize GCAM-USA v6 which focuses on 50 states plus the District of Columbia and simulates sub-national economy and energy systems. GCAM-USA v6 incorporates socioeconomic drivers, energy transformation trends, and final energy services at the state level. It utilizes the Shared Socioeconomic Pathway-2 (SSP2) growth assumptions \cite{o2017roads}. GCAM-USA v6 incorporates updated electricity technology cost assumptions from the National Renewable Energy Laboratory's (NREL) 2022 Annual Technology Baseline (ATB) \cite{EIA}. Transportation cost and energy intensity assumptions, including those for EVs, are primarily based on NREL's Electrification Futures Study \cite{jadun2017electrification}.

\subsection{Thermodynamic Global Warming Simulations (TGW)}
\label{sec:wrf}

The Thermodynamic Global Warming (TGW) simulations \cite{tgw-wrf, tgw-wrf1} are based on the Weather Research and Forecasting (WRF) model \cite{wrf, Skamarock2019}. WRF is a widely used numerical weather prediction model that calculates atmospheric variables at horizontal and vertical grid cells above the Earth's surface \cite{wrf, Skamarock2019}. The TGW simulations provide hourly and three-hourly meteorological data at a 12 km$^2$ resolution, covering the conterminous U.S. and parts of Canada and Mexico. For 2020-2099, the TGW simulations replay historical weather events under different levels of global warming based on multiple Representative Concentration Pathways (RCPs) and Global Climate Models. Warming levels are derived from average temperature and humidity changes from models that are ``cooler'' and ``hotter''  compared to the multi-model mean. We use four future climate scenarios obtained by combining the cooler and hotter scenarios with RCP 4.5 and RCP 8.5. 

\subsection{Total ELectricity Loads Model (TELL)}
\label{sec:tell}


The Total Electricity Loads (TELL) model \cite{McGrath2022} downscales annual state-level electricity demand projections from GCAM-USA to a hourly resolution. TELL utilizes multilayer perceptron models trained on historical hourly electricity demands and meteorological variations for each BA. The hourly BA-level loads from TELL are scaled to match the annual state-level loads from GCAM-USA, providing hourly non-transportation electricity demands for each BA. Combining these with transportation charging demands allows for a comprehensive analysis of change in total loads over time and across decarbonization scenarios.

\subsection{EVI-Pro Lite}
\label{sec:evi_pro}

EVI-Pro Lite \cite{center2020electric} is a data-driven tool that projects the aggregated charging demand of EVs. It uses detailed data on travel patterns, EV attributes, and charging infrastructure characteristics to generate charging electric load profiles. The tool is based on advanced PEV simulations trained on real-world driving data from large U.S. travel databases. Inputs to the tool include fleet size, discrete daily mean temperature, EV distribution, battery size, charging access, charging preferences (home vs work), and charging levels. We acknowledge that future technological advancements (e.g., wireless charging), changes in charging behavior, and broader EV adoption may challenge EVI-Pro assumptions based on historical data. Thus, we select parameters to represent realistic future charging scenarios. We detail our input assumptions and their relation with GCAM-USA, TGW, and TELL in Section \ref{sec:ldv_profile_method}.

\subsection{Fleet DNA Data}
\label{sec:fleetdna}

The Fleet DNA dataset \cite{fleetDNA} contains anonymous data on the mobility of commercial MHDVs in the U.S., including delivery vans, transit buses, and refuse trucks. We utilize specific data headings, such as "Vid" (vehicle identifier), "start\textunderscore{ts}" (vehicle record start time), "end\textunderscore{ts}" (vehicle record end time), and "distance\textunderscore{total}" (distance traveled by vehicle). To focus on return-to-base schedules, we exclude mobility instances that are less likely to be return-to-base. Also, we filter the data to include only vehicles that return to the depot by midnight and charge during their dwelling time in depots.

\subsection{Non-Road Vehicles Data}
Non-road vehicles data includes enplanements at U.S. commercial airports published by the U.S. Federal Aviation Administration \cite{enplanemnet}, route miles of rails, and shipping docks in U.S. published by the U.S. Department of Transportation \cite{rails_county_data}.


\section{Methodology}

\begin{figure}[!t]
    \centering
\includegraphics[width=0.99\columnwidth, clip=true, trim= 0mm 1.1mm 0mm 0mm]{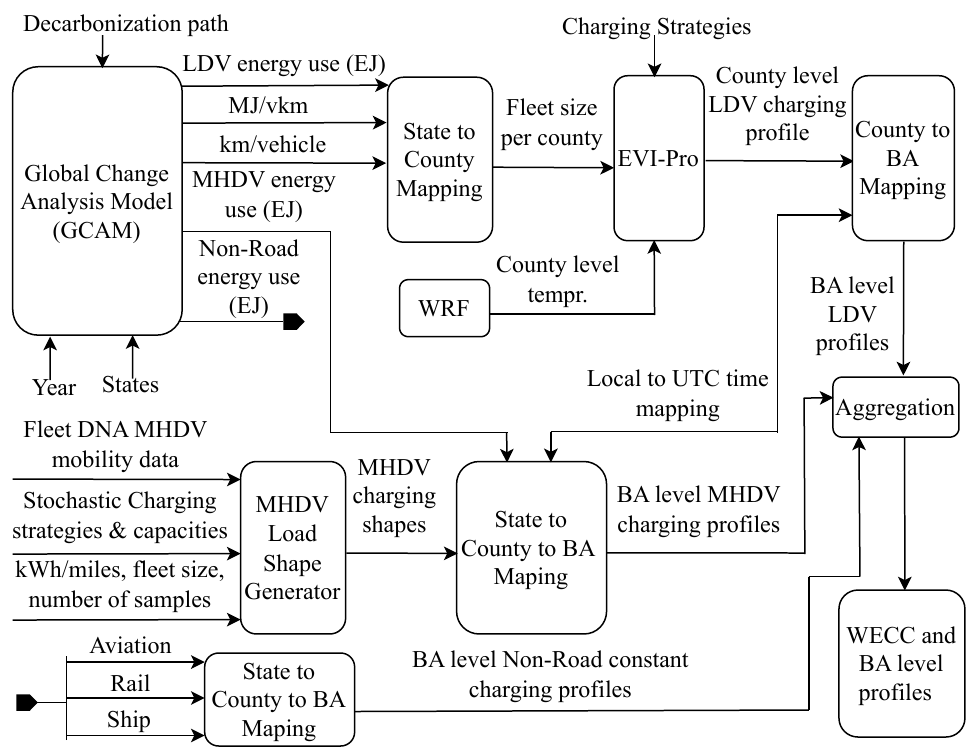}
\vspace{-15pt}
    \caption{Workflow for generating transportation charging load profiles across BAs in the western U.S. interconnection. Tempr.: Temperature, WRF: Weather Research and Forecasting Model, TELL: Total ELectricity Loads Model. }
    \label{fig:workflow}
\end{figure}

Fig.~\ref{fig:workflow} illustrates our methodology for developing spatio-temporal charging load profiles across BAs in the western U.S. interconnection. Below, we detail our approach.

\subsection{Developing LDV Charging Load Profiles}
\label{sec:ldv_profile_method}

Fig.~\ref{fig:workflow} illustrates the generation of LDV charging load profiles. This section outlines the methodology for translating the state-level annual energy consumption of electrified LDVs in GCAM-USA into hourly load shapes at the BA level.
\subsubsection*{Spatial Downscaling}
We use county-level resolution data to spatially downscale LDV charging load profiles. This allows us to incorporate temperature data from the TGW simulations (Section~\ref{sec:wrf}) and consider the diverse climatic regions within the BAs. The state-to-county distribution of energy use is determined by analyzing the distribution of registered EVs across counties using vehicle distribution data from 2018-2021. Each county's proportion of the state's total EVs is calculated, following the methodology outlined in \cite{kintner2020electric}.
\subsubsection*{Temporal Downscaling}
The county-level aggregate charging load profiles require various inputs from different modules. One crucial input for the EVI-Pro module is the fleet size. While direct fleet size data is not available from GCAM-USA, the annual energy consumption (in PJ/yr), energy usage per vehicle travel distance (in MJ/kvm), and average travel distance are utilized to estimate the fleet size. Daily mean temperature from TGW simulations (Section \ref{sec:wrf}) is used in EVI-Pro, with mapping to the nearest permissible discrete values. Additional inputs for EVI-Pro are outlined in Table \ref{T:evi_input}, independent of other modules in this study.

While EVI-Pro \cite{center2020electric} does not cover all future charging scenarios, we outline our key parameter choices in the model.
\begin{itemize}
    \item Battery EVs have a 250-mile range- the max. battery size in EVI-Pro. Plug-in hybrid EVs are excluded in this study.

    \item Predominant use of level-2 chargers (208-240 V) over level-1 chargers (120 V AC).
    
    \item Anticipated decrease in home charging preference from 80\% to 60\%, influenced by increasing chargers in public and work. Also, as EV ownership reaches 75-90\%, chargers in multi-family residential dwellers increases, reducing home charging preferences. In 2022, $\approx 63\%$ of houses are single-family units \cite{homeowner}, which also bounds the preference for home charging.
    \item Study of two charging strategies: \emph{min\_delay} for immediate maximum-speed charging after arrival and \emph{load\_level} for slow charging during the dwelling time. By 2035, 30\% of LDVs are expected to use load leveling, which is projected to increase to 70\% by 2050. Managed charging with price incentives and larger battery capacities reduce range anxiety, increasing load-leveling adoption.
\end{itemize}

\subsection{Developing MHDV Charging Load Profiles} 
\label{sec:mhdv_profile_method}

Fig.~\ref{fig:workflow} illustrates the two-step process for generating MHDV charging load profiles. 

In Step 1, using the MHDV load shape generator in Fig.\ref{fig:workflow}, we generate separate normalized MHDV charging load profiles for MDVs and HDVs. Inputs to the generator include mobility data for MHDVs (delivery vans, delivery trucks, school buses, transit buses, bucket trucks, tractors, and refuse trucks) obtained from Fleet DNA (Section\ref{sec:fleetdna}), MHDV charging strategies, EVCS capacities (\unit[]{kW}), unit energy use (\unit[]{kWh/miles}), fleet size, and number of sample fleets \cite{borlaug2021heavy}.
Three charging strategies—immediate, delay, and constant minimum power—are utilized. Immediate charging begins upon depot arrival and continues until the battery is fully recharged or the next trip commences. Delayed charging delays charging to ensure full recharge just before the subsequent trip. Minimum power charging involves charging at a constant minimum power level throughout the depot dwelling period to guarantee a full recharge for the next trip. We incorporate the uncertainty of future EV charging, considering a mix of charging strategies and multiple charging capacities, by assigning weights to both the charging strategies and charger capacities. The unit energy use (\unit[]{kWh/mi}) of vehicles depends on their weight, which is determined using a survey \cite{smith2020medium}. To minimize bias in MHDV profiles, we simulate the MHDV load shape generator (Fig.~\ref{fig:workflow}) for a user-defined fleet size and number of fleet samples, following the approach in \cite{borlaug2021heavy}. The load shape generator in Fig.\ref{fig:workflow} aggregates Fleet DNA mobility data by month and generates normalized average daily MHDV charging shapes with a 1-hour resolution for each month. These daily shapes are extended to create normalized yearly MHDV charging load profiles. Due to insufficient mobility data, we assume that the daily MHDV charging shape remains consistent throughout a month.

In Step 2, we scale the normalized MHDV charging load shapes by the annual MHDV energy use (in \unit[]{EJ}) projected by GCAM-USA in the western U.S. interconnection for a given year and decarbonization pathway. The scaled MHDV charging profiles in the western U.S. interconnection are further downscaled to BAs by considering the relative penetration of MHDVs across BAs. The penetration of MDVs in some BAs is reported in \cite{kintner2020electric}. For the remaining BAs we extrapolate the penetration of MDVs using the penetration of LDVs as a reference. Additionally, we assume that the relative penetration of MDVs and HDVs is the same across BAs.

\begin{table}[!t]
\centering
\caption{Choice of input parameter for EVI-Pro}
\vspace{-2pt}
\label{T:evi_input}
\vspace{-1mm}
\resizebox{1\columnwidth}{!}{
\begin{threeparttable}
\begin{tabular}{|c|c|c|c|}
\hline
 \rule{0pt}{2ex}\textbf{PEV}& BEV250 & \textbf{Class} & Equal\\ \hline 
 \textbf{Preference} & Home60 &
\textbf{Home access} & HA75, HA100 \\ \hline

 \rule{0pt}{2ex}\textbf{Home power} & MostL2 & \textbf{Home ch. strategy} & min\_delay, load\_leveling \\ \hline
 
 \rule{0pt}{2ex}\textbf{Work power} &MostL2  &\textbf{Work ch. strategy} & min\_delay \\ \hline
\end{tabular}
\end{threeparttable}
}
\end{table}

\vspace{-11pt}
\subsection{Developing Non-Road Vehicles Charging Load 
Profiles}
\label{sec:non-road}

As Fig.\ref{fig:workflow} shows, we generate constant charging profiles for non-road vehicles (aviation, rails, and ships). To downscale the state-wise yearly energy use by non-road vehicles in GCAM-USA to hourly charging power across BAs in the western U.S. interconnection, we use the ratio of i) airport enplanements for aviation, ii) route miles travelled in railroads and transits for rails and trains, and iii) shipping docks for ships, in a given BA to that of the entire western U.S. interconnection. The energy at the western U.S. interconnection level is determined by aggregating the state-level energies in GCAM-USA for the 11 states within the region. In this study, we assume that electrified rails, ship, and aviation consume constant power over each hour of the year. We make this assumption due to the lack of data on their charging behavior as they are not yet widely deployed. Therefore, this study may not reflect accurate temporal downscaling of non-road vehicles. However, their spatial downscaling is based on county-level data.

\section{Case Study and Results}
This section examines transportation load profiles in the western U.S. interconnection for 2035 and 2050, considering different decarbonization pathways and climate scenarios. It compares the sensitivity of these profiles and projected loads to the total system load. We present profiles for NZ decarbonization, highlighting its aggressiveness compared to BAU pathway. Our open-source code allows for profile generation in 5-year increments until 2050 for both BAU and NZ pathways.

\subsection{Sensitivity of LDV Load Profiles}
\label{sec:ldv_sensitivity}
\begin{figure}[!t]
\centering
\subfigure[\label{fig:sensitivity_charging_strategy_mdv}]{
\includegraphics[width=0.31\columnwidth, clip=true, trim= 2.5mm 0mm 10mm 5mm]{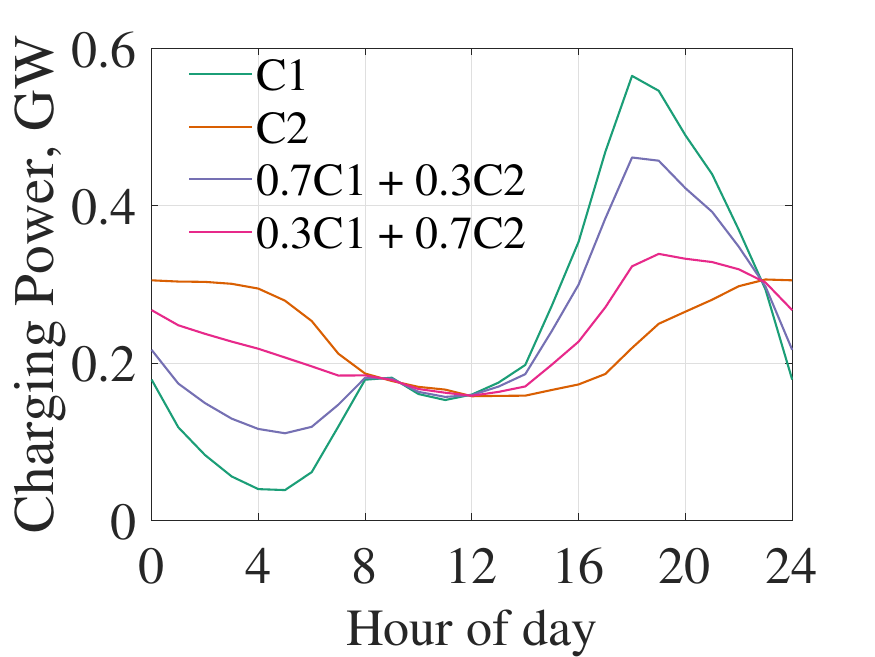}}
\hfill
\subfigure[\label{fig:sensitivity_charger_mdv}]{
\includegraphics[width=0.31\columnwidth, clip=true, trim= 2.5mm 0mm 10mm 5mm]{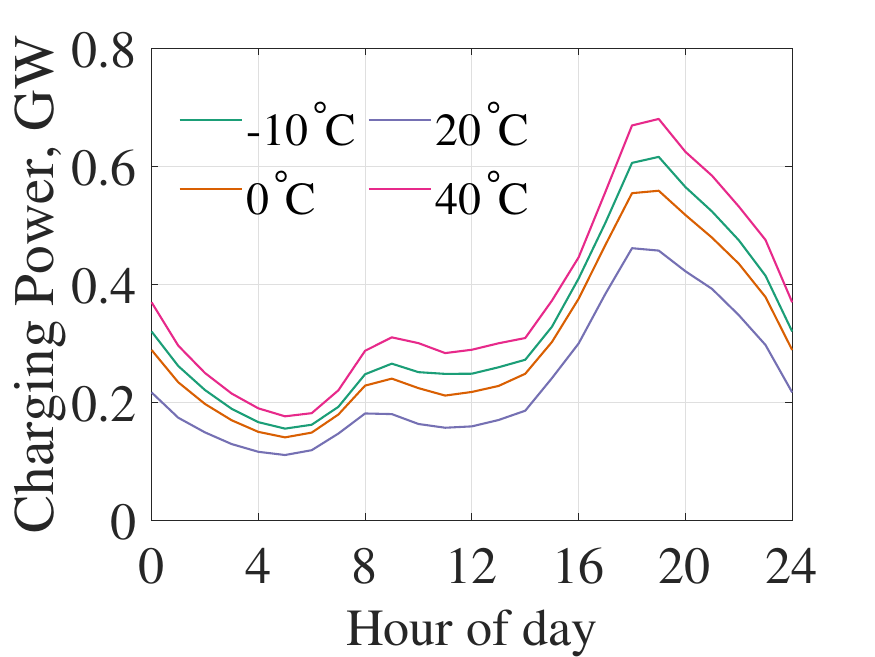}}
\hfill
\subfigure[\label{fig:sensitivity_n_samples_mdv}]{
\includegraphics[width=0.31\columnwidth, clip=true, trim= 1mm 0mm 8mm 5mm]{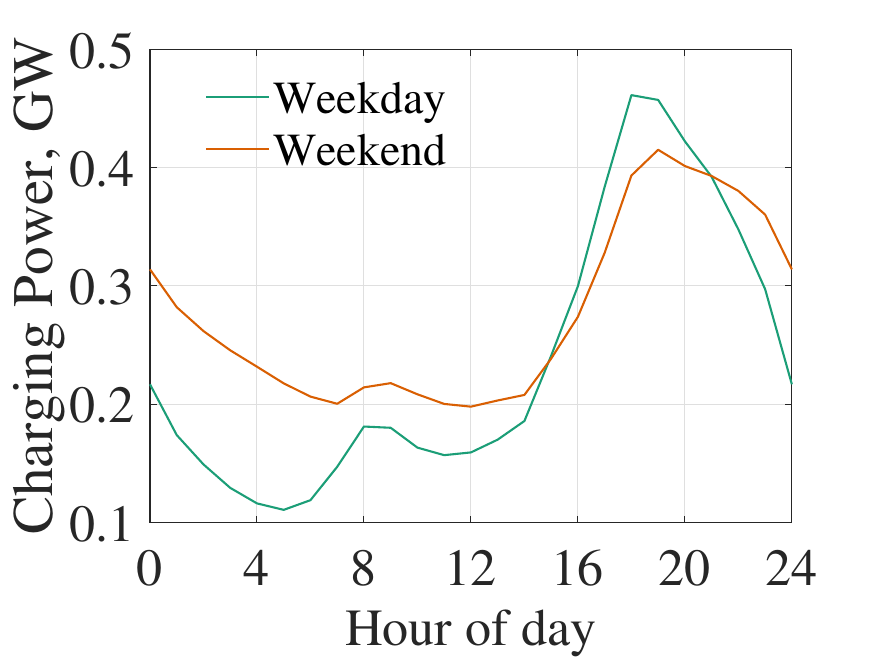}}
\caption{Sensitivity of LDV average daily charging load profiles (00:00 to 24:00 hours in PDT) to (a) charging strategies, (b) temperature, and (c) weekends and weekdays. In (a), C1 and C2 are immediate (\emph{min\_delay}) and load levelling (\emph{load\_level}) charging strategies. The results in (a)-(c) are simulated for King County, Washington in 2035 using the NZ decarbonization pathway. }
\label{fig:sensitivity:ldv}
\end{figure}

Fig. \ref{fig:sensitivity:ldv} demonstrates the impact of LDV charging strategies, temperature, and weekdays vs. weekends on the LDV charging profiles. In Fig. \ref{fig:sensitivity:ldv}(a), load-leveling charging  (C2) shows a flat, grid-friendly charging profile compared to immediate charging (C1). With a 70\% adoption rate of C2 by 2050, the peak load can be reduced by 40\% compared to the C1-only charging. Fig. \ref{fig:sensitivity:ldv}(b) shows the impact of temperature on LDV charging. A temperature of $20 ^\circ$C requires the least charging, while $-10$, $0$, and $40^\circ$C  days consume  $28\%$, $21\%$, and $36\%$ more power, respectively. Temperature affects charging efficiency and energy consumption for the same travel distance. Fig. \ref{fig:sensitivity:ldv}(c) shows LDV charging peak load during weekends is approximately 10\% lower than on weekdays.

\subsection{Sensitivity of MHDV Load Profiles}
\label{sec:mhdv_sensitivity}
\begin{figure}[!t]
\centering
\subfigure[\label{fig:sensitivity_charging_strategy_mdv}]{
\includegraphics[width=0.31\columnwidth, clip=true, trim= 2.5mm 0mm 10mm 5mm]{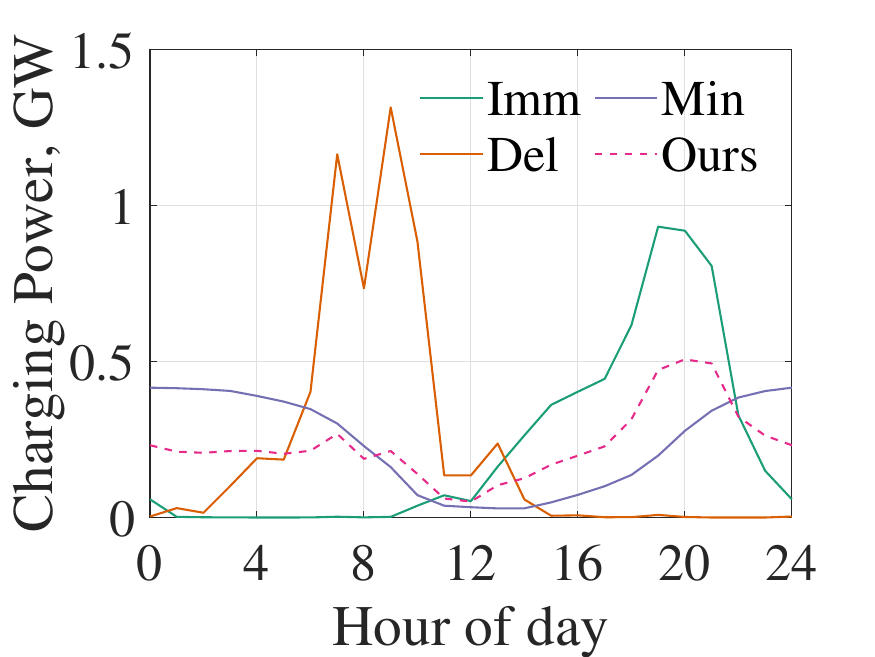}}
\hfill
\subfigure[\label{fig:sensitivity_charger_mdv}]{
\includegraphics[width=0.31\columnwidth, clip=true, trim= 2.5mm 0mm 10mm 5mm]{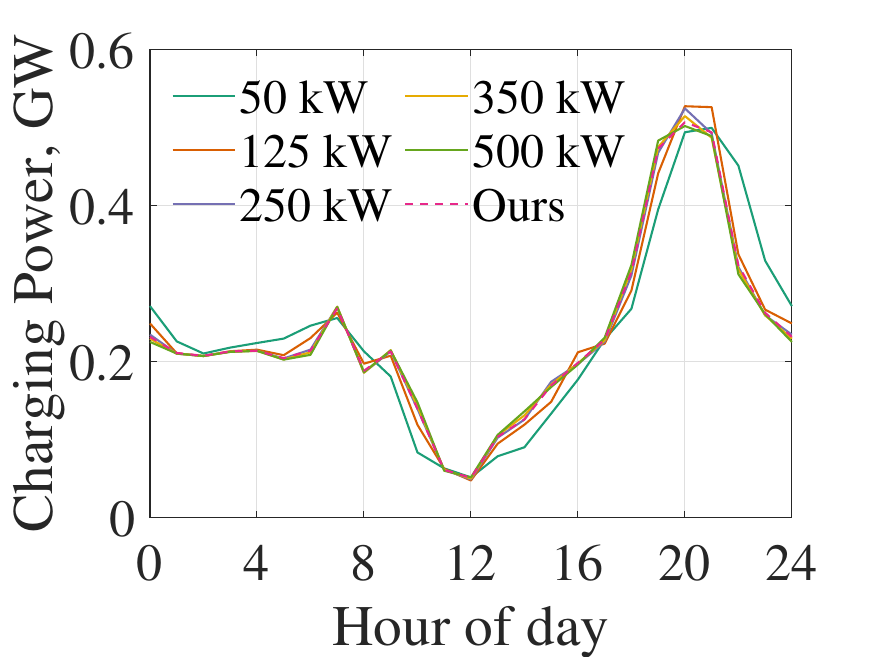}}
\hfill
\subfigure[\label{fig:sensitivity_n_samples_mdv}]{
\includegraphics[width=0.31\columnwidth, clip=true, trim= 1mm 0mm 8mm 5mm]{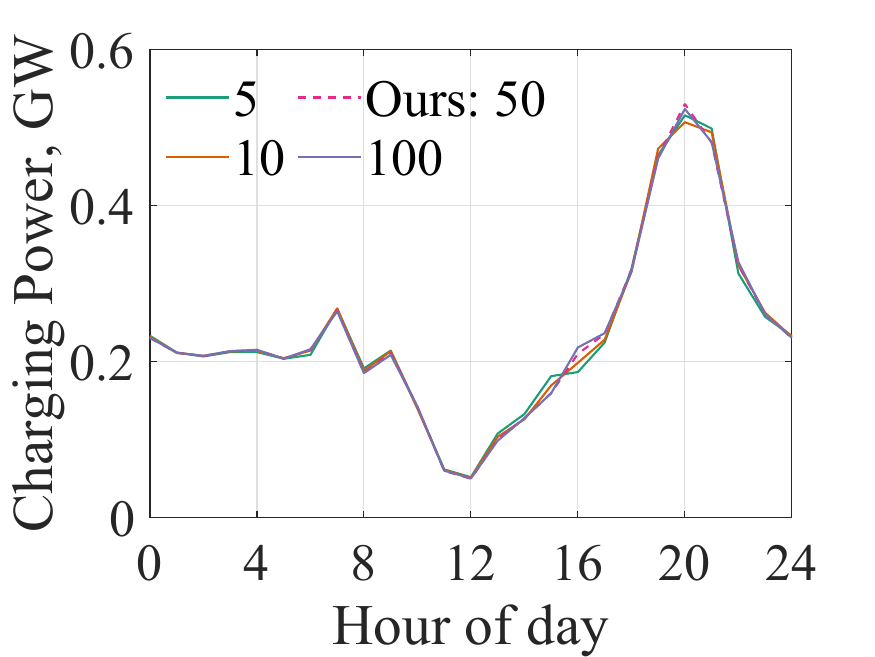}}

\subfigure[\label{fig:sensitivity_charging_strategy_hdv}]{
\includegraphics[width=0.31\columnwidth, clip=true, trim= 2mm 0mm 10mm 5mm]{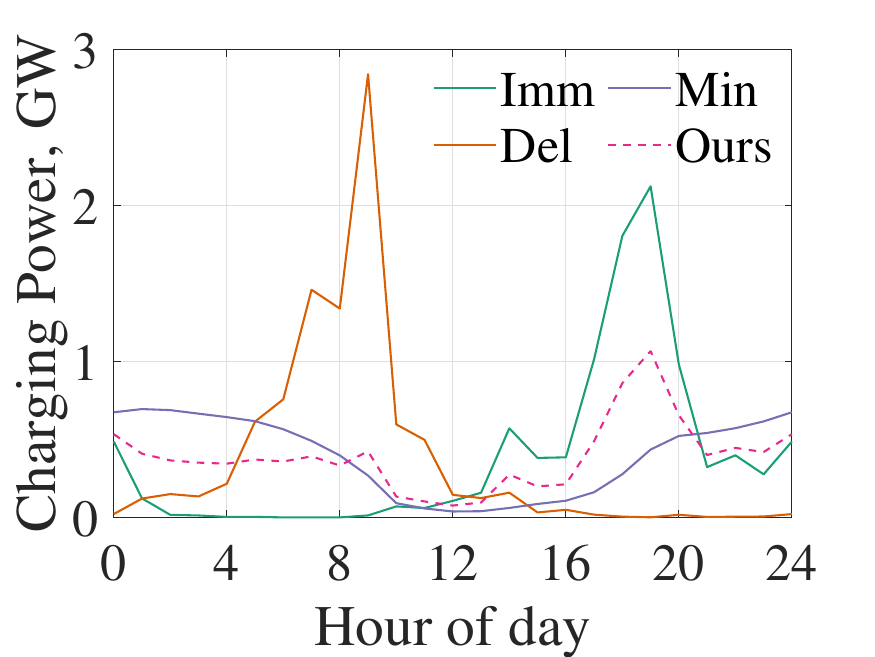}}
\hfill
\subfigure[\label{fig:sensitivity_charger_hdv}]{
\includegraphics[width=0.31\columnwidth, clip=true, trim= 2mm 0mm 10mm 5mm]{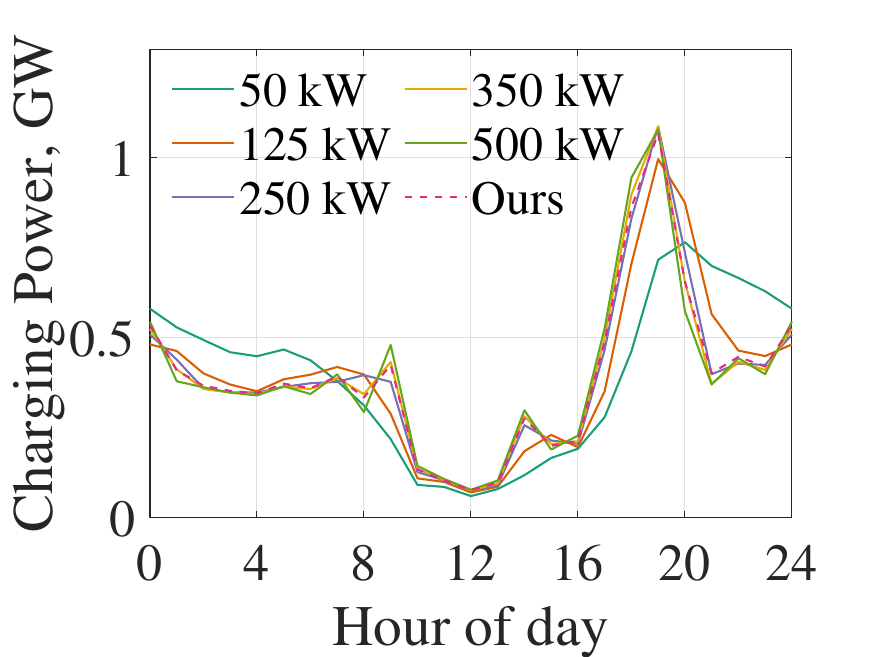}}
\hfill
\subfigure[\label{fig:sensitivity_n_samples_hdv}]{
\includegraphics[width=0.31\columnwidth, clip=true, trim= 2mm 0mm 10mm 5mm]{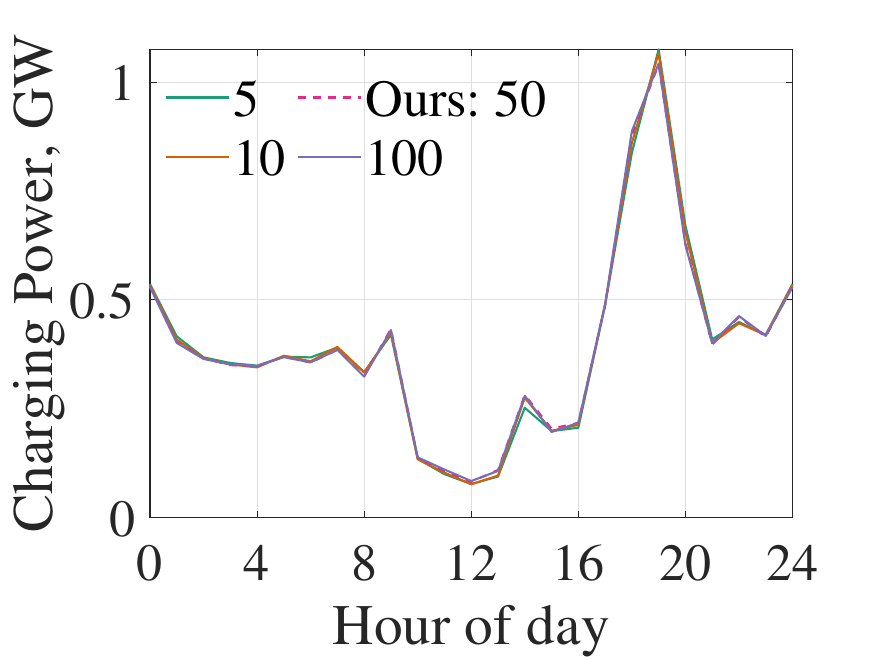}}
\caption{Sensitivity of MHDV average daily charging profiles (00:00 to 24:00 hours in PDT), (a)-(c): MDV and (d)-(f): HDV, with (a) and (d): charging strategies, (b) and (e): charger capacities, and  (c) and (f): number of sample fleets for stochastic simulation. In (a) and (d), Imm: Immediate charging, Del: Delayed charging, Min: Minimum power charging, Ours: 40\% Imm + 10\% Del + 50\% Min strategies. In (b) and (e), Ours: 5\% 50 kW + 5\% 125 kW + 10\% 250 kW + 40\% 350 kW + 40\% 500 kW chargers. The profiles in (a)-(f) are simulated for the CISO BA in 2035 using the BAU pathway.}
\label{fig:sensitivity:mhdv}
\end{figure}

Fig.~\ref{fig:sensitivity:mhdv} shows the sensitivity of average daily MHDV charging load profiles to charging strategies, charger capacities, and number of sample fleets for stochastic simulation. As Figs.~\ref{fig:sensitivity_charging_strategy_mdv} and \ref{fig:sensitivity_charging_strategy_hdv} show, the choice of charging strategy significantly impacts charging power's magnitude and timing, resulting in varying charging peaks. For instance, as Fig.~\ref{fig:sensitivity_charging_strategy_hdv} shows, the HDV peak is 0.67 GW for minimum power charging at 00:00 while it is 2.84 GW for delay charging at 9:00; representing a 323\% increase. To achieve a more balanced and grid-friendly charging profile, we mix 40\% immediate, 10\% delay, and 50\% constant minimum power charging strategies as an example. This mix means majority of MHDVs charge with constant power throughout their dwelling time or start charging upon arrival at depots and a small portion wait for a suitable time to charge, such as during periods of cheaper electricity prices, before their departure. Charger capacities also influence the charging profiles, with slower charging leading to smaller peaks. For example, as Fig.\ref{fig:sensitivity_charger_hdv} shows, the 50 kW charger exhibits a peak of 0.76 GW at 20:00, while fast chargers have a peak of 1.08 GW at 19:00, representing a 42\% increase. We anticipate that larger battery capacities will drive the adoption of high power chargers. Thus, we consider a mix of chargers: 5\% 50 kW, 5\% 125 kW, 10\% 250 kW, 40\% 350 kW, and 40\% 500 kW. Furthermore, as Figs.~\ref{fig:sensitivity_n_samples_mdv} and \ref{fig:sensitivity_n_samples_hdv} show, the charging profiles remain stable across different fleet samples, despite variations in the number of sample fleets.

\subsection{Transportation Load Profiles}
\label{sec:load_profiles}

\begin{figure}[!t]
\centering
\includegraphics[width=0.97\columnwidth, clip=true, trim= 10mm 0mm 12mm 2mm]{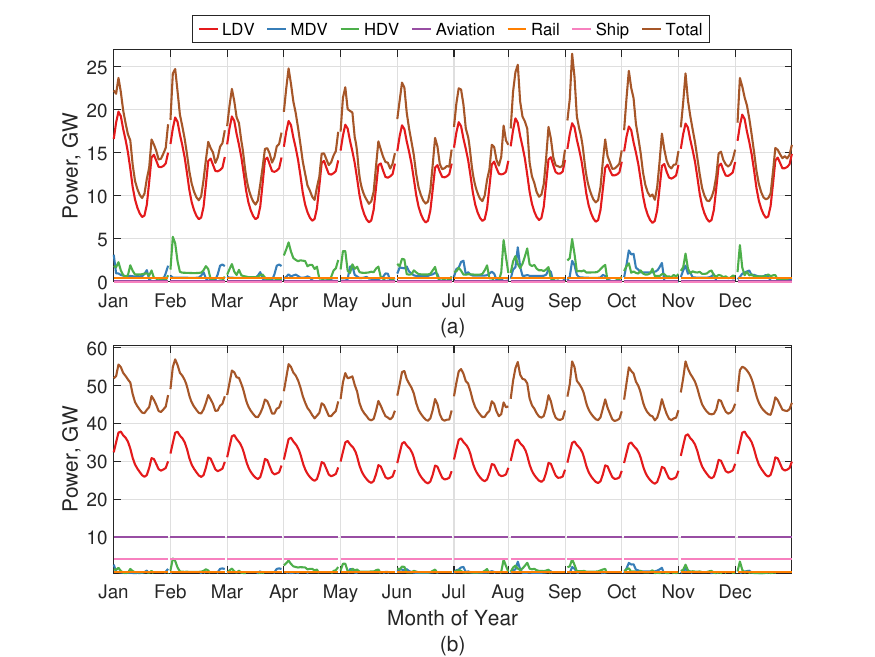}
\vspace{-10pt}
\caption{Average daily transportation charging load profiles (00:00 to 23:00 UTC hours) in the western U.S. interconnection across months in (a) 2035 and (b) 2050 for the NZ decarbonization pathway and RCP 4.5 climate scenario.}
\label{fig:profiles_NZ}
\end{figure}

Figs.~\ref{fig:profiles_NZ}(a) and (b) depict the  charging load profiles (00:00 to 23:00 UTC hours discontinuous for each month) for LDVs, MHDVs, rails, aviation, ship, and total transportation in the western U.S. interconnection in 2035 and 2050 under the NZ decarbonization pathway. Several key dynamics are observed. Firstly, the charging load profiles differ by vehicle type. Secondly, the transportation peak load significantly increases from 27.77 GW in 2035 to 59.85 GW in 2050, marking a 115\% surge. Additionally, the variation in charging load experiences a 6\% increase, with values of 19.19 GW in 2035 and 20.35 GW in 2050. Thirdly, LDVs contribute prominently to the total transportation charging load profile. Fourthly, charging load for non-road transportation modes (aviation, rail, and ship) exhibits an upward trend from 2035 to 2050. Finally, the MHDV charging load slightly decreases in 2050 compared to 2035 (e.g., HDV peak in 2035 is 4.9 GW but 4.1 GW in 2050) due to the NZ decarbonization pathway's emphasis on transitioning from MHDVs, particularly MDVs, to mass transportation options such as rail and ships. This decline is further driven by the adoption of alternative clean fuel technologies, such as hydrogen, in later years.

\subsection{Transportation Electric Load Relative to System Load}
\label{sec:load_ratio}

Three metrics (M1, M2, and M3) in Fig.~\ref{fig:load_ratio} assess the spatial and temporal impact of transportation load on the electric power system. M1 represents the yearly ratio of total electric energy to transportation electric energy, M2 measures the ratio of transportation load to system load at the system peak, and M3 measures the ratio of transportation load to system load at the transportation peak. M1 rapidly increases in 2050 relative to 2035, indicating steep increase in electrification across all BAs. The results also reveal that all metrics vary significantly across BAs. For instance, southern California BAs such as LDWPD and IID have high M1 (e.g., in IID the M1 is 0.50 in 2035 and 0.68 in 2050). This is attributed to the aggressive EV adoption trend in the region and the predominantly residential system load with a moderate climate. Similarly, CHPD in Chelan County, Washington, has high M1 due to its low non-transportation electric load and ambitious EV policy. Notably, the transportation peak and system peak do not coincide, reducing system stress from transportation charging.

\subsection{Load Profiles with BAU Decarbonization}
\begin{figure}[!t]
\centering
\includegraphics[width=0.97\columnwidth, clip=true, trim= 12mm 0mm 14mm 4mm]{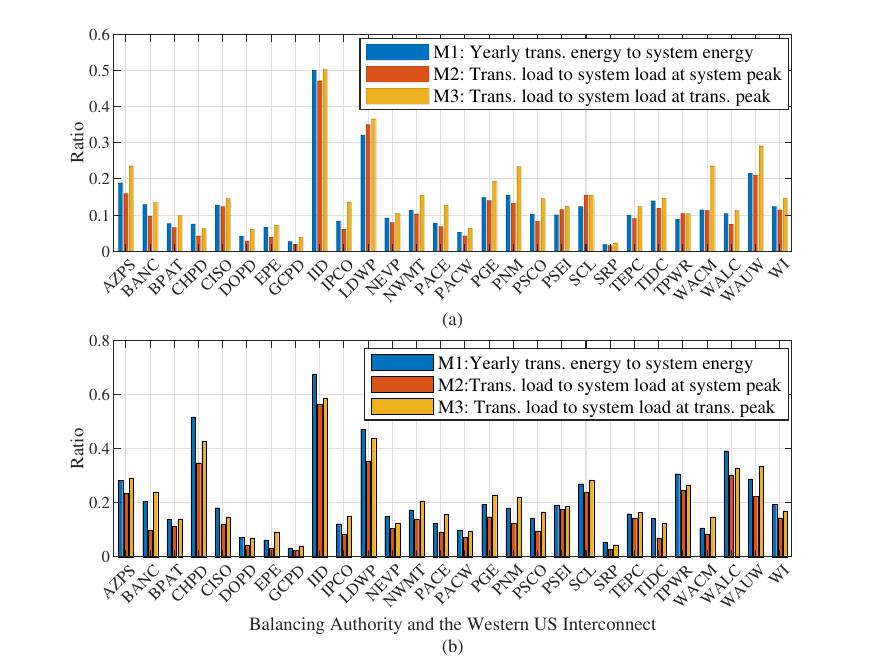}
\caption{Comparison of electric transportation (trans.) to system load for the western U.S. interconnection (WI) BAs in (a) 2035 (b) 2050 for the NZ decarbonization pathway and RCP 4.5 climate scenario.}
\vspace{-15pt}
\label{fig:load_ratio}
\end{figure}

In comparison to the BAU pathway (not shown in this paper), the NZ pathway (Figs.~\ref{fig:profiles_NZ} and \ref{fig:load_ratio}) demonstrates significant differences in charging peak load and variation. In 2035, NZ has a 4.7\% higher charging peak load than BAU (27.77 GW vs. 26.51 GW), increasing to approximately 65.4\% in 2050 (59.85 GW vs. 36.17 GW). NZ also exhibits higher load variation, with about 21.5\% more variation in 2035 (19.19 GW vs. 15.79 GW) and 10.5\% more variation in 2050 (20.35 GW vs. 18.41 GW) compared to BAU. Additionally, the transportation-to-system energy ratio is notably higher in the NZ pathway, particularly in 2050. For example, in IID, NZ shows a 9.8\% higher ratio than BAU in 2050 (0.67 in Fig.~\ref{fig:load_ratio}(b) vs. 0.61).

\section{Conclusion}
 

This paper presents a novel approach for generating spatially-distributed hourly time-series of transportation charging load profiles. Applied to the western U.S. interconnection in 2035 and 2050, our analysis reveals that transportation charging loads can contribute significantly to the system electric peak, ranging from 2.4\% to 56.6\% in different BAs, despite accounting for less than 20\% and 17\% of total electric load for the 2050 NZ case and the 2050 BAU case, respectively. This variation is influenced by factors such as load nature, climate zones, and EV adoption trends. Understanding this non-uniform spatial and temporal impact is crucial for effective decarbonization and transportation electrification policies. To facilitate further research and policy analysis, we provide a publicly available dataset and code at \url{https://doi.org/10.5281/zenodo.7888569}. Researchers and decision-makers can use this resource to develop charging load profiles for other regions, enabling informed decision-making in the pursuit of sustainable transportation electrification.

\section*{Acknowledgment}
This research was supported by the GODEEEP Investment at Pacific Northwest National Laboratory (PNNL). PNNL is a multi-program national laboratory operated for the U.S. Department of Energy (DOE) by Battelle Memorial Institute under Contract No. DE-AC05-76RL01830. We thank Kate Forrest and Brian Tarroja at University of California, Irvine, USA for the discussions on the transportation profiles.

\vspace{-11pt}
\bibliographystyle{IEEEtran}
\bibliography{ref}
\end{document}